\documentclass[10pt,english,aps,amssymb,prb,twocolumn, showpacs, superscriptaddress]{revtex4-2}
\pdfoutput=1
\usepackage[T1]{fontenc}
\usepackage[latin9]{inputenc}
\usepackage{amsmath}
\usepackage{graphicx}
\usepackage{amssymb}
\usepackage{esint}
\usepackage{esint}
\usepackage{verbatim}
\usepackage{xcolor}

\makeatletter
\@ifundefined{textcolor}{}
{% \definecolor{BLACK}{gray}{0}
 \definecolor{WHITE}{gray}{1}
 \definecolor{RED}{rgb}{1,0,0}
 \definecolor{GREEN}{rgb}{0,1,0}
 \definecolor{BLUE}{rgb}{0,0,1}
 \definecolor{CYAN}{cmyk}{1,0,0,0}
 \definecolor{MAGENTA}{cmyk}{0,1,0,0}
 \definecolor{YELLOW}{cmyk}{0,0,1,0}
 }

\newcommand{\bra}[1]{\langle #1|}
\newcommand{\ket}[1]{|#1\rangle}

\renewcommand{\phi}{\varphi}
\renewcommand{\epsilon}{\varepsilon}

\newcommand{\ii}{\mathrm{i}}

\usepackage{babel}

\begin{document}

\title {Topological Phases on Quantum Trees}
\author{Alex Weststr\"om}
\affiliation{Department of Physics, School of Science, Westlake University, Hangzhou, Zhejiang, China}
\affiliation{Key Laboratory for Quantum Materials of Zhejiang Province, School of Science, Westlake University, Hangzhou, Zhejiang, China}
\affiliation{Institute of Natural Sciences, Westlake Institute for Advanced Study, Hangzhou, Zhejiang, China}
%\email[Correspondence to ]{}
\author{Wenbu Duan}
\affiliation{Department of Physics, School of Science, Westlake University, Hangzhou, Zhejiang, China}
\affiliation{Key Laboratory for Quantum Materials of Zhejiang Province, School of Science, Westlake University, Hangzhou, Zhejiang, China}
\affiliation{Institute of Natural Sciences, Westlake Institute for Advanced Study, Hangzhou, Zhejiang, China}
\author{Kangpei Yao}
\affiliation{College of Electrical and Information Engineering, Hunan University, Changsha, Hunan, PR China}
\author{Xiaonan Wang}
\affiliation{College of Electrical and Information Engineering, Hunan University, Changsha, Hunan, PR China}
\author{Jie Liu}
\affiliation{College of Electrical and Information Engineering, Hunan University, Changsha, Hunan, PR China}
\author{Jian Li}
\affiliation{Department of Physics, School of Science, Westlake University, Hangzhou, Zhejiang, China}
\affiliation{Key Laboratory for Quantum Materials of Zhejiang Province, School of Science, Westlake University, Hangzhou, Zhejiang, China}
\affiliation{Institute of Natural Sciences, Westlake Institute for Advanced Study, Hangzhou, Zhejiang, China}
\date{\today}
\begin{abstract}

In this work, we present a theory for topological phases for quantum systems on tree graphs. Conventionally, topological phases of matter have been studied in regular lattices, but also in quasicrystals and amorphous settings. We consider specific generalizations of regular tree graphs, and explore their topological properties. Unlike conventional systems, infinite quantum trees are not finite-dimensional, allowing for novel phenomena. We find a proliferation of topological zero modes present throughout the entire system, indicating that the bulk also acts as a boundary. We then go on to show that only three symmetry classes host stable topological phases in contrast to the usual five symmetry classes per dimension. Finally, we introduce what we call the Su-Schrieffer-Heeger tree which is topologically non-trivial even in the absence of inner degrees of freedom and does not possess any gapped trivial phases. We realize this system in an electronic circuit and show that our theory matches with experiments.
       
\end{abstract}
%\pacs{73.63.Nm, 73.63.-b, 03.65.Vf}
\maketitle
\bigskip{}

\section{Introduction}

Topology has in recent years become one of the main pillars upon which the theoretical foundation of condensed matter physics rests. For non-interacting symmetry-protected topological phases, one of the most integral parts of the theory is the periodic table of topological phases \cite{kitaev:2009, schnyder:2009, ryu:2010}. By knowing the presence or absence of the three symmetries (assuming no other symmetries present) time-reversal, charge-conjugation and chiral symmetry together with the dimension of a quantum system, the table can immediately tell us how many distinct gapped topological phases can in principle be found in such a quantum system. For each spatial dimension, there are always five out of the ten symmetry classes which have a non-trivial classification. 

\begin{figure}
	\includegraphics[width=\linewidth]{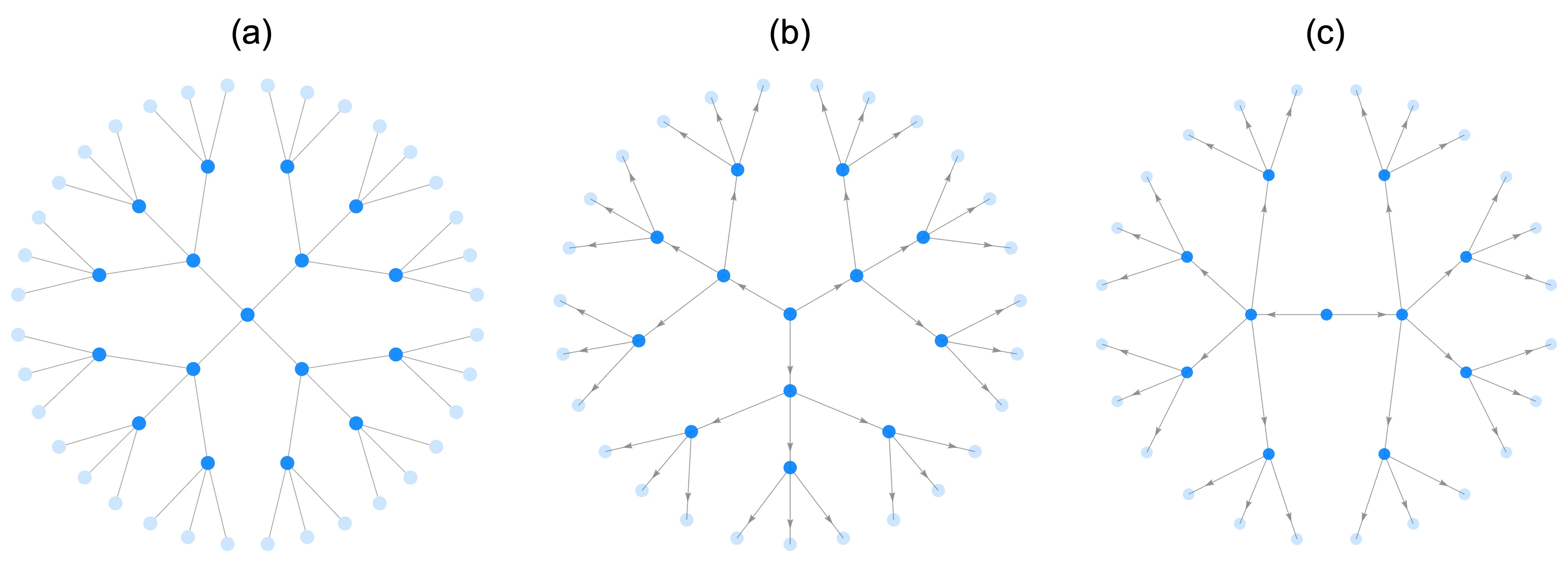}
	\caption{Graphical representations of different tree structures: (a) a conventional Bethe lattice with coordination number $z=4$; (b) an RQT with constant branching number $n=3$; (c) an arbitrary RQT.}
	\label{fig:betherqt}
\end{figure}

Furthermore, the periodic nature of the table allows us to consider systems in arbitrarily many spatial dimensions which seemingly covers all possible cases one could think of, but is it possible to device a system which is topologically non-trivial yet does not directly fall into the periodic table?

One possibility to consider is including additional symmetries. This approach is perhaps the most prominent one and has lead to the discoveries of concepts such as topological crystalline insulators and higher-order topological phases \cite{fu:2011, schindler:2018}. Another alternative is to allow for non-Hermitian Hamiltonians, which extends the number of possible symmetry classes and brings a whole host of new features to the table \cite{bernard:2002, esaki:2011, gong:2018, liu:2019, wojcik:2020}.
% However, the work in this direction has largely been reliant on the (integer-valued) dimension of the space into which the fractal has been embedded. 

%A fourth possibility is to consider systems with infinite spatial dimensionality. In \textbf{REF}, they considered a Chern insulator in infinite dimensions by considering a sequence of even-dimensional systems and taking the limit as the number of dimensions go to infinity. However, the sequence of finite-dimensional systems always has the same classification so there is never any ambiguity in the infinite limit.

A third possibility is to study systems which are inherently infinite-dimensional. One such system is the Bethe lattice (BL) which is a type of infinite tree graph. The BL was originally used within statistical physics \cite{bethe:1935}. More generally, tree graphs emerge in physics in the context of Feynman diagrams when calculating the contributions of the lowest-order (in $\hbar$) processes to the path integral. Outside of physics tree graphs naturally pop up whenever there are hierarchical structures present (e.g. databases).

After first fixing a coordination number $z$, the BL is constructed sequentially by first introducing one node -- also known as the root node -- which we say to be the zeroth generation. The first generation is then generated by introducing $z$ neighbors to the first node. All subsequent generations are then given by attaching $z-1$ new neighbors to all the nodes in the previous generation. Note that there are no closed loops in this setup such that the BL is nothing but an infinite Cayley tree. An example of a BL with $z=4$ is depicted in Fig.~\ref{fig:betherqt}(a).
%One way to see how the BL is infinite-dimensional is to consider the number of nodes $N(R)$  within a radius $R$ from any given node. For any (finite) $d$-dimensional system, we have $N(R)\sim R^d$. However, for the BL, the relation is exponential $N(R) \approx z^R$ which grows faster than any finite polynomial.

Some work related to BLs and other types of topology have already been done \cite{depenbrock:2013, sen:2020, manoj:2021} but so far topological phases of non-interacting Fermions have remained unexplored; hence, in this work, we present a theory of topological phases on what we dub rooted quantum trees (RQTs) which generalize the BL in a couple of ways as described in the paragraph below.

In \cite{mahan:2001, petrova:2016, yorikawa:2018, aryal:2020}, a band theory on the simplest nearest-neighbor scalar hopping model sans any inner degrees of freedom on a BL was presented. A band-theory formulation often makes the analysis of topological phases straightforward, but for our purposes, we wish to extend this theory to systems with inner degrees of freedom. In such systems, we no longer necessarily have hopping terms that satisfy $\hat t = \hat t^\dag$. As such, viewing the original BL as an undirected graph where the hopping terms indicate the connectivity (i.e., edges) between nodes, it is then natural to assign a directionality to the edges by taking the direction of the edge to correspond to the hopping direction associated with the matrix $\hat t$. Similarly, Hermiticity dictates that hopping against the direction (or arrow) of the edge is associated with the matrix $\hat t^\dag$. With this directionality in place, we can now conclude that the same undirected BL can be the platform to many different inequivalent directed graphs. For RQTs, the arrows between two generations of nodes all point from the lower generation to the higher as seen in Fig.~\ref{fig:betherqt}(b). Furthermore, while the hopping terms between generations $g$ and $g+1$ are all the same $\hat t_g$, we still allow for the possibility that $\hat t_g \neq \hat t_{g^\prime}$ when $g\neq g^\prime$. In addition to this, we also allow the coordination number to be generation-dependent $z\to z_g$. An example of such a more general RQT is depicted in Fig.~\ref{fig:betherqt}(c). Here, the coordination numbers can be seen to be $z_0=2$, $z_1=5$, and $z_2=4$. As is evident from the figure, the root node is a natural choice for the origin in an RQT, in contrast to the conventional BL where all nodes are equivalent. 

The content of the rest of the manuscript is as follows: in Sec.~\ref{sec:topprop}, we introduce the theory of topological phases on the RQT. We show how the distinction between the bulk and the boundary of the RQT becomes blurred where the direct consequence is that the topological phase of an RQT comes equipped with topological zero modes present throughout the whole system. We further argue why only three symmetry classes can host topologically non-trivial phases as opposed to the conventional five per spatial dimension. We illustrate our conclusions using numerical calculations on an RQT which we dub the Kitaev tree. In Sec.~\ref{sec:ssh}, we present the so-called Su-Schrieffer-Heeger tree, which is a topologically non-trivial RQT without inner degrees of freedom and no trivial gapped phases. We implement this tree in an electrical circuit and show how the measurements match the predictions of our theory. Finally, in Sec.~\ref{sec:disc}, we summarize our results and provide a brief outlook on future directions.

\section{Topological Properties}\label{sec:topprop}

In order to understand the topological properties of RQTs, we decompose the infinite-dimensional system to something finite dimensional and more familiar. This decoupling process is explained in detail in App.~\ref{app:decomp}, but the basic idea is to exploit the permutation symmetry that exists at every generation of the RQT: by construction, the $z_g-1$ higher-generation neighbors of any node in the RQT are the root nodes of identical sub-RQTs, meaning that changing their order in any way leaves the system invariant. This rich symmetry allows us to map the system to a set of decoupled linear chains as an isotypic decomposition in mathematical language (see App.~\ref{app:iso_dec}), which begin at some generation and extend all the way to the highest generation in the system -- note that this linear-chain picture works even if we restrict ourselves to a finite system with a finite number of generations. The number of chains starting at any given generation $g$ is given by $N_g - N_{g-1}$, where $N_g$ denotes the number of physical sites in generation $g$. In terms of the coordination numbers, they are $N_0 =1$, $N_1 = z_0$, and $N_g = z_0\prod_{j=1}^{g-1}(z_j-1) = \prod_{j=0}^{g-1}n_j$ for $g>1$. Here, we have introduced the branching number $n_g$ which is defined to be $n_0 = 1$, $n_1 = z_0$ and $n_g = z_{g-1}-1$ for $g>1$.

The bulk part of the Schr\"odinger Equation for a linear chain starting at generation $G$ is given by
\begin{equation}\label{eq:schreq}
	\begin{split}
		\varepsilon\left.\left\vert G, g\right.\right\rangle &= \hat h_{G+g}\left.\left\vert G, g\right.\right\rangle\\
		&+ \sqrt{n_{G+g}} \hat t_{G+g} \left.\left\vert G, g + 1\right.\right\rangle\\
		&+ \sqrt{n_{G+g-1}}\hat{t}_{G+g-1}^\dag \left.\left\vert G, g - 1\right.\right\rangle.
	\end{split}
\end{equation}
Here $\left.\left\vert G, g\right.\right\rangle$ is a state in generation $G + g$ ($g\geq 0$) in the set of states comprising a chain starting at generation $G$. As mentioned, the number of states starting at a given generation is generally larger than one, but for the sake of clarity, we have omitted adding an additional chain index in the above equation. The matrices $\hat h_g$ and $\hat t_g$ are the onsite and hopping terms, respectively. Evidently, the coordination numbers renormalize the hopping terms within the chains. Consider now chains starting at generations $G$ and $G^\prime$ and take $g\gg G,G^\prime$. Far from the edge of either chain, the bulk looks the same for both and given the finite correlation lengths in the system, we expect degeneracies to arise not only from equivalent states in different chains starting at the same generation, but also from equivalent bulk states living in chains starting at different generations when the system is infinite.  The boundary condition is given by
\begin{equation}
	\varepsilon\left.\left\vert G, 0\right.\right\rangle = \hat h_{G}\left.\left\vert G, 0\right.\right\rangle + \sqrt{n_{G}} \hat t_G \left.\left\vert G, 1\right.\right\rangle.
\end{equation}

Given the freedom to choose the onsite and hopping terms as well as the coordination numbers, we now see that we can realize \textit{any} one-dimensional nearest-neighbor model. Consequently, to make our RQT topologically non-trivial, we apparently only need to pick our parameters such that the chains correspond to one-dimensional topologically non-trivial phases. While this is true (although with a certain caveat as we will explore later on), it does not mean that the topology of RQTs is completely equivalent to that of one-dimensional systems, but the linear-chain picture does provide a bridge between them.

One of the most prominent novel features is that of the \textit{bulk-boundary duality} (BBD) in RQTs: as is already hinted at by the fact that there are chains starting at every generation, there is a sense in which the bulk also contains edges. The consequence is that once the RQT is in a non-trivial phase, there is a proliferation of zero-modes throughout the whole system. In a pristine RQT, the BBD can be considered a direct consequence of the conventional bulk-boundary correspondence in one-dimensional systems. 

However, it is then important to ask how the linear-chain pictures holds up in the presence of disorder -- indeed, to this end, we must first answer an even bigger question which is whether or not the phases are stable under disorder; while it is true that an individual chain is stable as per conventional wisdom, the RQT is a set of chains that once we introduce disorder couple to each other. This may in turn lead to the coupling and gapping out of zero modes of different chains. In fact, this is precisely what we expect in the two symmetry classes which have a $\mathbb{Z}_2$ classification in one dimension, i.e. D and DIII. In contrast, the remaining non-trivial classes AIII, BDI, and CII can be expected to be robust even when the chains couple. 

To be more explicit, we first consider the case where we have an infinite clean RQT in class AIII: the edge modes for each chain will have the same chirality by virtue of having the same bulk equation Eq.~\eqref{eq:schreq}. Since a state with definite chirality cannot have a non-zero energy, and any arbitrary superposition of states of the same chirality still maintains the same chirality, the zero modes remain at zero even when we turn on the disorder. 

The story for BDI and CII is similar to that of AIII, but since we have both time-reversal and charge-conjugation symmetry in these systems, we must make sure that they do not map to states of different chirality (both symmetries are local, so if they map to states with opposite chirality, we know we have states of both chiralities living in the same physical location). In the presence of time-reversal ($\mathcal T$) and charge-conjugation ($\mathcal{P}$), the chiral operator is given by $\mathcal{C} = \mathcal{TP}$. Since the chiral operator squares to unity, we have $\mathcal{TP} = \mathcal{P}^{-1}\mathcal{T}^{-1}$ which implies that $\mathcal{T}$ and $\mathcal{P}$ commute only if they both square to the same value (either $1$ or $-1$). If they commute, they also commute with the chiral operator and hence do not map between states of different chiralities. 

From this, we also see that in the case of DIII, the two operators do map to opposite chiralities, but if there are only two states at the edge, they cannot be removed from zero since the spinful time-reversal enforces a degeneracy, which is why their classification is $\mathbb{Z}_2$ rather than $\mathbb{Z}$. 

Finally, for class D, there is only particle-hole symmetry with the symmetry operator squaring to unity. If there is one edge mode at zero, it cannot be gapped out, since that would require it to split into two separate states that move symmetrically away from zero energy. 

In summary, all this gives us the second novel feature of RQTs: there are only three symmetry classes (AIII, BDI, and CII) with non-trivial topology in RQTs. In finite-dimensional systems, there are always five symmetry classes with non-trivial classifications for each dimension.

To illustrate the above conclusion, we will consider a ``Kitaev tree'' with uniform coordination number $z$, where the bulk equation~\eqref{eq:schreq} is that of Kitaev's $p$-wave superconductor \cite{kitaev:2001}. The onsite term is $\hat h = -\mu \sigma^y$, and the hopping is $\hat t = t\sigma^y - i\Delta\sigma^x$ with $\mu$, $t$, and $\Delta$ being the chemical potential, the hopping amplitude, and superconducting pairing, respectively. As usual, $\sigma^{x,y,z}$ are the Pauli matrices. This system has a chiral symmetry effected by the operator $\mathcal{C}=\sigma^z$, plus time-reversal and charge-conjugation symmetry corresponding to the operators $\mathcal{T} = \sigma^z K$ and $\mathcal{P} = K$, respectively. Here, $K$ denotes complex conjugation. Without further context, these symmetries put the system squarely into class BDI. However, if we introduce a pairing disorder $\delta\hat t_{ij} = i\sigma^z \tau_{ij}$, where $i,j$ are the indices of two neighbors, and $\tau_{ij}$ is antisymmetric and sampled from some random distribution, we break the time-reversal symmetry, and change the class to D. Based on what we know from the familiar one-dimensional model, our system is topologically non-trivial whenever $\vert \mu\vert < 2\sqrt{z-1}\vert t\vert$. Below, we will first consider an onsite disorder which does not break any of the symmetries and then compare that with that of the time-reversal-symmetry breaking $\delta\hat t_{ij}$.

\begin{figure}
	\includegraphics[width=\linewidth]{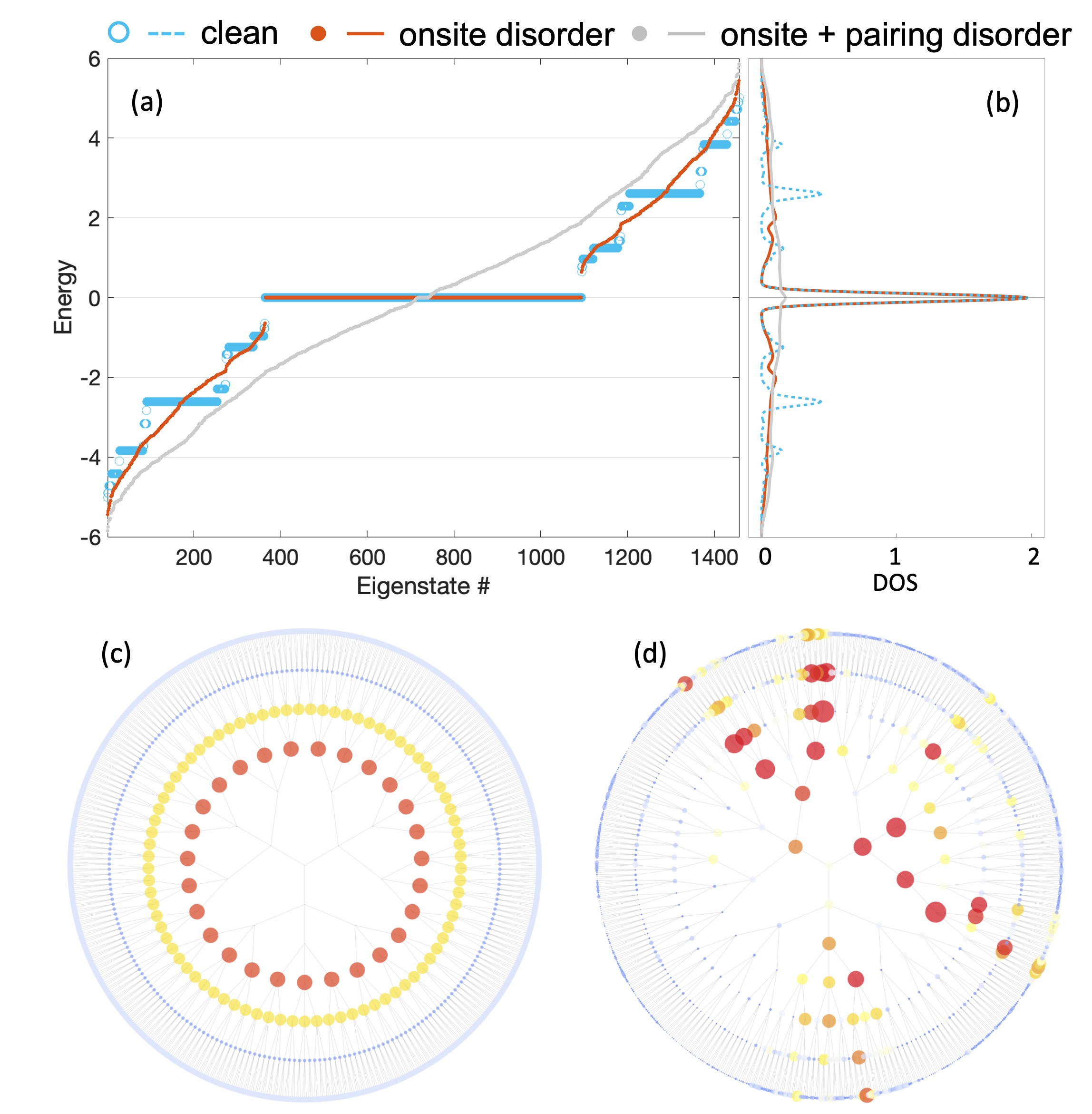}
	\caption{Eigenstates in clean and disordered Kitaev tree. (a) Spectra of Kitaev trees in the clean limit (cyan) and with disorder that preserves (red) or breaks (gray) the chiral symmetry. (b) the density of states (DOS) corresponding to the spectra shown in (a), broadened slightly for a better view. (c) and (d) Probablity weights of a typical zero-energy eigenstate (c) in the clean limit or (d) with chiral-symmetry-preserving disorder, where the value of the weight on each node is indicated by the color and the size of the marker. The parameters used here are: $z = 4$, $t = 1$, $\mu = 1.6t$ and $\Delta = 0.1$. The onsite and the pairing disorder strength parameters (where applicable) are $20\sqrt{z-1}\Delta$ and $3\sqrt{z-1}\Delta$, respectively. The DOS in panel (b) is plotted with a broadening bandwidth being $0.1$, and an average of $1000$ samples in each disordered setting.}
	\label{fig:kitaev}
\end{figure}

From the numerical results presented in Fig.~\ref{fig:kitaev}, we can see that the zero modes persist even in the presence of a large onsite disorder while a relatively weak pairing disorder begins to gap them out. We can also further see from Fig.~\ref{fig:kitaev}(a) that the degeneracies of the bulk states in the pristine system break down along with the linear-chain picture itself; while the BBD was initially argued from the chain picture, the persistence of the topological zero modes (in the absence of the symmetry-breaking pairing disorder) tells us that this feature is not merely a quirk of clean RQTs. This can be seen more clearly from the density of states depicted in Fig.~\ref{fig:kitaev}(b). The clean system has peaks corresponding to all eigenenergies of the system, whereas all peaks save for the one arising from the topological zero-modes are washed out in the presence of onsite disorder. 

From the BBD, we expect zero-energy states to be spread throughout the entire system. In Fig.~\ref{fig:kitaev}(c), we see a representative zero-energy state located in the central region of the system. Had we summed over the amplitudes for all of the zero-energy states, we would see a more or less uniform amplitude on all nodes with any deviations being attributable to finite-size effects. Similarly, in Fig.~\ref{fig:kitaev}(d), we plot a representative for a system with onsite disorder. The state can still be seen to be spread out over the system, albeit with less symmetry than for the clean system.  Of course, in an infinite and therefore perfectly degenerate system, one could cook up asymmetric zero-modes by appropriate linear combinations of the rotationally symmetric (in terms of their amplitudes) states in the clean system. It might be tempting to somehow attribute the BBD to the existence of an origin in RQTs. However, as we show in App.~\ref{app:rootless}, we can construct RQTs \textit{without} root nodes, where there are no preferred nodes nor natural origins.

To conclude this section, we now give some brief comments on how the numerical calculations were done in practice: we have used the self-energy of the system to emulate an infinite system in a finite number of nodes. In practice, this means that we have included large imaginary terms at the highest generation of the finite system. These imaginary terms are chosen such that they respect the symmetry of the system. More specifically, for the spectra and wavefunctions shown in Fig.~\ref{fig:kitaev}, we have used the self-energy (at zero energy) obtained from the numerically calculated boundary Green's function for the decoupled chains of the clean RQT. In the cases where disorder is introduced, the self-energy further includes a number (chosen to be 2 for the data shown in Fig.~\ref{fig:kitaev}) of equally disordered terminal sites in the virtual chains to ensure there are no spurious zero modes remaining at the highest generation.

%\begin{enumerate}
%	\item Linear Chain
%		\begin{itemize}
%			\item Group theory
%			\item New chains
%			\item Renormalized hopping
%			\item Schreq
%			\item Any NN model
%		\end{itemize}
%	\item Bulk-Edge
%		\begin{itemize}
%			\item Chains starting everywhere
%			\item topology $\to$ ZM everywhere
%			\item Stable against disorder, bulk is not
%			\item rootless RQT
%		\end{itemize}
%	\item Stability
%		\begin{itemize}
%			\item Disorder couples chains
%			\item $Z_2$
%			\item Stability argument
%			\item Only three stable classes
%			\item Kitaev
%			\item D class disorder
%			\item \textbf{$n_G - 2n_{G-1}$ versus $n_G - n_{G-1} + n_{G-1}$}
%		\end{itemize}
%\end{enumerate}

\section{The Su-Schrieffer-Heeger Tree in a Topoelectric Circuit}\label{sec:ssh}

In the previous section, we saw how we could construct topologically interesting phases in RQTs by a suitable choice of inner degrees of freedom. In this section we present the so called Su-Schrieffer-Heeger (SSH) tree, where we forgo any inner degrees of freedom and consider the simplest possible hopping model on an RQT. However, the twist is now that we will employ the physical structure itself to encode effective degrees of freedom by allowing the coordination number to alternate between generations. Since the coordination number enters Eq.~\eqref{eq:schreq} as renormalization factors to the hopping terms, the chains are now formally equivalent to the famous SSH model \cite{su:1979}. The theory of the SSH model is already well-established and we can immediately conclude three things: first, no matter which two coordination numbers we pick, as long as they are not equal, there will be topological zero modes in the system. Second, these modes will only be present on every second generation. Third, since the zero modes are a consequence of having two alternating branching numbers, there is no way to tune the system into a trivial phase without either altering the physical structure itself, or making it gapless -- there are no gapped trivial phases. 

Owing to its simplicity, it is straightforward to construct a real-world version of the SSH tree using electrical circuits. The framework for simulating topological quantum systems in this way was put forth in \cite{lee:2018}. As explained in App.~\ref{app:topcirc}, there exists an immediate relationship between the wavefunctions and energies of the quantum system, and the impedance measurements of the corresponding electrical circuit given by
\begin{equation}
	Z_{ab} = \sum_{n} \frac{\Vert \psi_{n,a} - \psi_{n,b}\Vert^2}{j_n}.
\end{equation}
Here, $Z_{ab}$ is the impedance between nodes $a$ and $b$, $\psi_{n,a}$ is the value of the $n$th wavefunction at node $a$, and $j_n$ is the $n$th eigenvalue of the circuit Laplacian, which is a function of both the frequency $\omega$ and the $n$th eigenvalue of the quantum system.

\begin{figure*}
	\includegraphics[width=\textwidth]{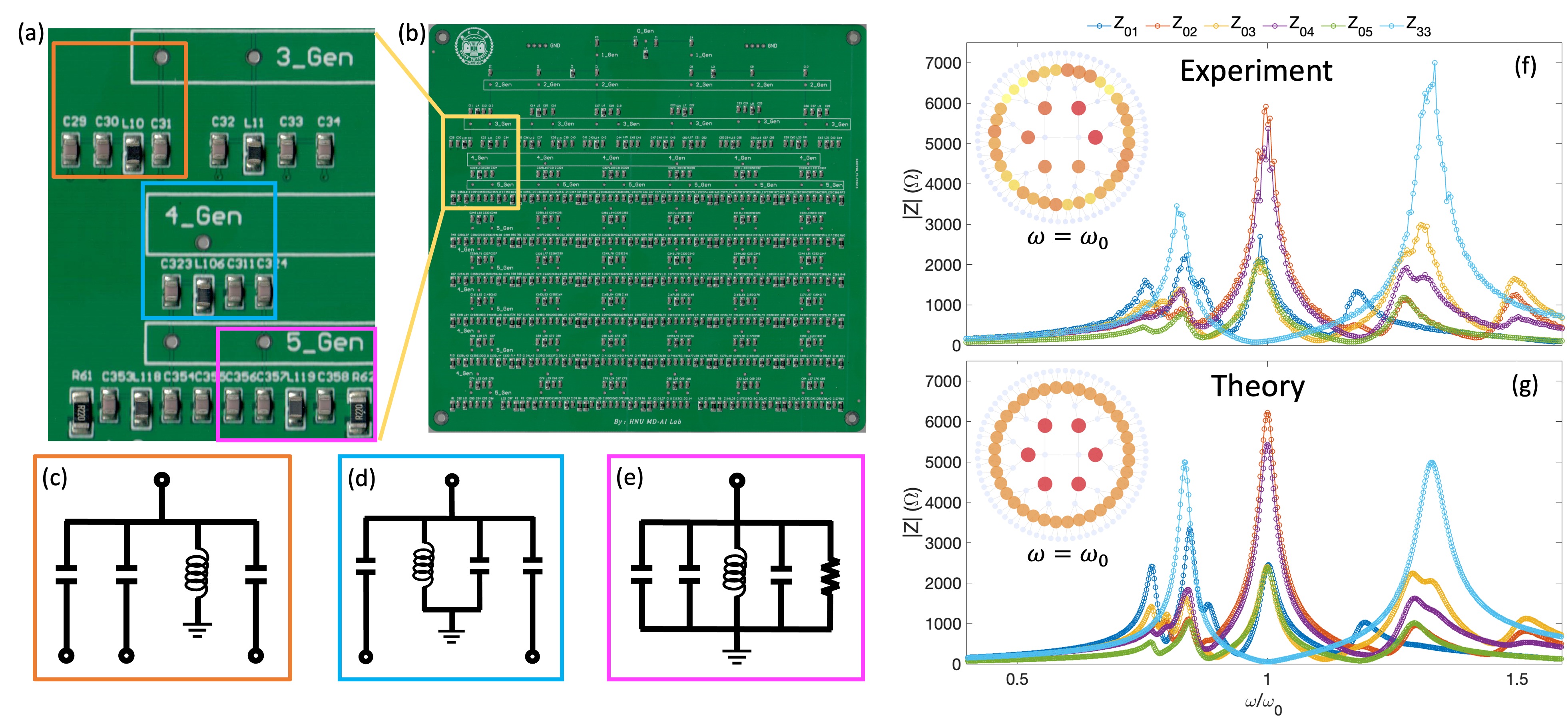}
	\caption{Electric circuit simulation of a Su-Schrieffer-Heeger Tree. (a) and (b) Photos of the electric circuit board at different scales. (c), (d) and (e) Circuit diagrams for three key building blocks (also highlighted in (a) with boxes of the same colors) that the full circuit is composed of. (f) and (g) Frequence-dependence of the magnitude of the impedance (f) measured in the experiment or (g) calculated with our theory, for six characteristic combinations of nodes. The presence of two types of peaks and one type of total suppression around $\omega=\omega_0$ (see the text) is a signature of topological zero modes. In the insets, the impedance magnitude maps at $\omega=\omega_0$, with one node fixed at the root and the other exhausting all remaining nodes, are also shown.}
	\label{fig:exp}
\end{figure*}

The hopping terms of the quantum system are all real and can without loss of generality be taken to be positive. In the electrical circuit, such terms will be effected by capacitors. The inter-capacitor leads then take the role of the sites. We further couple each of the sites to the ground via an inductor, but also a capacitor in parallel in some cases to ensure that each site couples to the same number of capacitors in total. The last generation of sites are further connected to the ground via resistors, which correspond to the non-Hermitian imaginary term in the quantum system that simulates an infinite system or a drain. The circuit along with schematic pictures of its components can be found in Fig.~\ref{fig:exp}(a) and (b), and (c-e), respectively. Its technical specifications can be found in App.~\ref{app:topcirc}.

In Fig.~\ref{fig:exp}(f) and (g), we see a comparison between the numerically calculated and the experimentally measured impedance over a range of frequencies between the root node and one typical node in each of the other generations. In the numerical calculation we have added small resistances in series to the capacitors as the leads in the physical circuit are not perfect. We see that the numerical results based on our SSH-tree model is in excellent agreement with the real-world data. The resonance peak at $\omega\approx 1.6\cdot 10^6 rad/s$ corresponds to the topological zero modes of the quantum system. Owing to their topological origin, this peak is robust against any chiral-symmetry-preserving imperfections of the circuit board and is present for all nodes in even-numbered generations.

\section{Discussion}\label{sec:disc}

In this work, we have studied topological phases on RQTs. Owing to their dimensionality, they fall outside of the periodic table of topological phases, yet has an intimate connection to one-dimensional topology. However, the connection is also not an equivalency. We demonstrated what we call the BBD of RQTs first in the clean limit with the linear-chain picture, but also through the persistence of robust topological zero modes even when disorder made the linear-chain picture invalid. Furthermore, we saw that unlike finite-dimensional systems, there are only three instead of five symmetry classes that can support non-trivial topological phases.

We also showed how effective inner degrees of freedom can be encoded into the physical structure itself. Particularly, the SSH tree has no inner degrees of freedom, yet is always in a non-trivial topological phase as long as the chiral symmetry and gap are maintained. To reiterate: there are no gapped trivial phases. In the end, we implemented the SSH tree in a topoelectric circuit and showed that the impedance measurements agree very well with our theory.

The RQT represents one of many possible quantum systems one can put on a tree graph. The symmetry of the RQT allows us to straightforwardly decompose it into linear chains from which its topological properties can be understood. It would be interesting to consider for example unrooted quantum trees, where each node has equally many arrows pointing inwards as outwards. This structure would no longer enjoy the same permutation symmetries as the RQT and would have to be solved in a wholly novel way.

\acknowledgments
This work is supported by the National Natural Science Foundation of China (\#92265201, \#61804049); the Fundamental Research Funds for the Central Universities of P.R. China; Huxiang High Level Talent Gathering Project (\#2019RS1023); the Key Research and Development Project of Hunan Province, P.R. China (\#2019GK2071); the Technology Innovation and Entrepreneurship Funds of Hunan Province, P.R. China (\#2019GK5029); the Fund for Distinguished Young Scholars of Changsha (\#kq1905012).

\appendix

\section{Decomposing the RQT}\label{app:decomp}

In this section, we provide a detailed account of how to decompose a RQT into linear chains. To this end, let us first introduce some notation. A physical site in generation $g$ can be labelled by a sequence of numbers $i_1,i_2,\ldots,i_{g}$, where each $i_g\in \mathbb{N}$ goes from $1$ to $n_{g}$. The root node is labelled by $0$. Then, the first generation nodes have the index $i_1$. A node in the second generation is labelled by two indices, $i_1$ and $i_2$. The first index tells us which of the first-generation nodes it connects to, and the second index labels which among the second-generation nodes connecting to the first-generation node $i_1$ it is. The extension to all higher generations follows the same pattern. We can then denote any quantum state living on site $i_1,i_2,\ldots,i_g$ by $\vert i_1 i_2\cdots i_{g}\rangle$. For the moment, let us assume we have a maximum number of generations $G$.

\subsection{Tree Symmetry Group}
Now, the RQT is invariant under arbitrary permutations of any branches at any generation in the system. For example, the first generation has $n_1$ sites which are each a root node for their own subtrees. We can permute the order of these trees however we want without changing the structure of the RQT. This tells us that it is symmetric under the symmetric group $S_{n_1}$. Similarly, we can permute all the subtrees of any one of these subtrees without changing the physical structure. Since they act independently on each set of subtrees, we have a direct sum of $n_1$ copies of the symmetric group $S_{n_2}$. We can extend this to all the subtrees starting from any generation to see that the \textit{tree symmetry group} $\mathcal G$ of the RQT is generated by considering all possible permutations of all these subtrees. Conversely, we can categorize subgroups of $\mathcal G$ in terms of which subtrees elements of a particular subgroup permute; a subgroup whose elements permute the $n_{g+1}$ subtrees whose root nodes lie in generation $g+1$ and connect to the same node in generation $g$, is said to act non-trivially on generation $g+1$.	

%The entire group is then generated by all these subgroups acting non-trivially on one generation only
%\begin{equation}
%	\mathcal G = \left\langle S_{n_1}, \bigoplus_{j_1=1}^{n_1}S_{n_2},\ldots,\bigoplus_{j_G=1}^{n_1 n_2 \cdots n_{G-1}}S_{n_{G}}\right\rangle,
%\end{equation}
%where $\oplus$ denotes a direct product of groups.

If $q_g\in \mathcal G$ denotes an element that only acts non-trivially on the $g$th generation, then the action of the group representation $\rho: \mathcal G\to GL(M, \mathbb{C})$ can be defined by
\begin{equation}\label{appeq:greponh}
	\rho(q_{g^\prime})\vert i_1\cdots i_{g}\rangle = \sum_{j_{g^\prime}=1}^{n_{g^\prime}} \rho(q_{g^\prime})_{i_{g^\prime}, j_{g^\prime}}^{i_1,\ldots,i_{g^\prime-1}}\vert i_1\cdots i_{j_{g^\prime}} \cdots i_{g}\rangle, 
\end{equation}
if $g^\prime \leq g$, otherwise it acts trivially. The upper indices on $\rho(q)$ are needed to distinguish its action on all the different branches in the given generation. Here, the $M$ in $GL(M, \mathbb{C})$ is the total number of sites in the system.

In the natural representation, the symmetric group $S_N$, $N\in \mathbb{N}$ consists of all the matrices which permute the elements of an $N$-dimensional complex vector space $\mathbb{C}^N$. This representation can be broken down into two irreducible representations (irreps): a trivial representation where one only considers the one-dimensional subspace where all coordinates are the same, and the so-called natural representation which is the complementary $N-1$-dimensional subspace where all coordinates sum to zero. The representation of the group $\mathcal G$ on our Hilbert space can also be broken down into irreps using the trivial and standard representations of individual symmetric groups.

First, we can identify a set of $G+1$ one-dimensional irreps of $\mathcal G$ in our Hilbert space through the vectors
\begin{equation}\label{appeq:trivialsub}
	\vert x, g\rangle \equiv \sum_{i_1,\ldots,i_{g}}x\vert i_1\cdots i_{g}\rangle,
\end{equation}
together with $x\vert 0\rangle$. Here $x\in \mathbb{C}$ is just some complex number. Each of these irreps consist of equal superpositions of all the sites in one generation. We can thus distinguish these irreps by which generation they belong to. For the irrep in generation $g$, the corresponding subspace is spanned by the unit vector given by normalizing the vector in Eq.~\eqref{appeq:trivialsub}.

Since the actions of the hopping term and its Hermitian conjugate in the RQT Hamiltonian on these states are
\begin{equation}\label{appeq:hop}
	\hat T \vert i_1\cdots i_{g}\rangle = \sum_{i_{g+1}=1}^{n_{g+1}} \hat t_{g-1}\vert i_1\cdots i_{g}i_{g+1}\rangle,
\end{equation}
and
\begin{equation}\label{appeq:hopherm}
	\hat T^\dag \vert i_1\cdots i_g\rangle = \hat t^\dag_g\vert i_1\cdots i_{g-1}\rangle,
\end{equation}
we can see that they will map between ``neighboring'' trivial irreps. From this we can conclude that the basis vectors of these irreps can be thought of as sites in a one-dimensional chain that extends from the root node all the way to the outermost generation. For reasons that should become clear in the following paragraph, we will denote an irrep living in generation $g$ in this sequence of trivial irreps by $(0, g)$.

\begin{figure}
	\includegraphics[width=\linewidth]{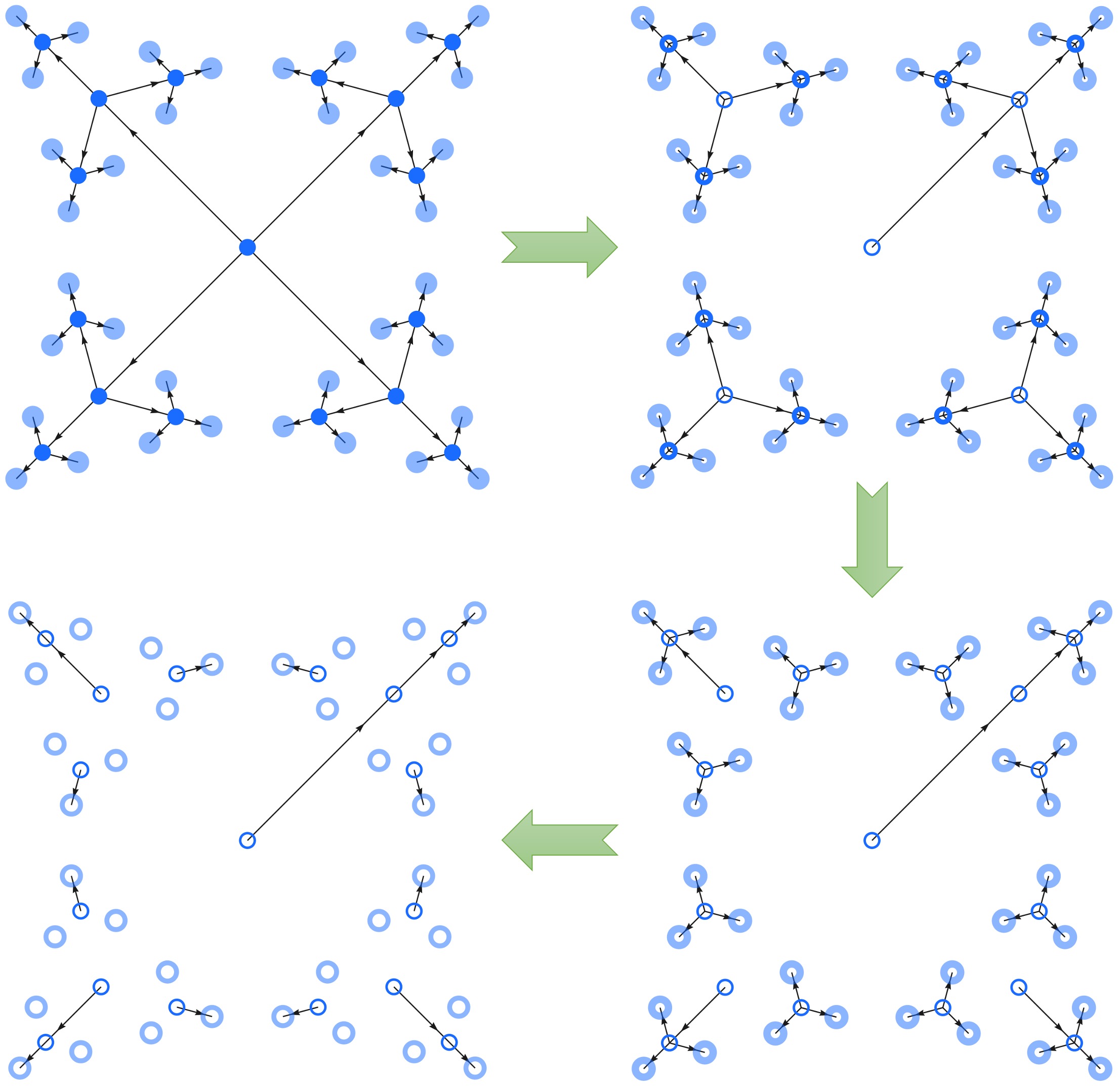}
	\caption{An illustration of the decoupling procedure, with guiding arrows, that transforms a symmetric RQT into a series of linear chains. In each step, a partial transformation is made with respect to the index for one of the generations, and this is done outwards from the first generation consecutively until the last one.}
	\label{fig:decoupling}
\end{figure}

Next we will find irreps that are related to the natural representation of the symmetric group. Each irrep can be labelled by two numbers $g$ and $g^\prime$ such that $1\leq g^\prime \leq g \leq G$ and a vector in the irrep $(g^\prime, g)$ can be expressed as
\begin{equation}\label{appeq:standardrepvec}
	\vert x_{g^\prime}, (g^\prime, g)\rangle = \sum_{i_1,\ldots,i_{g}}x^{i_1,\ldots,i_{g^{\prime}-1}}_{i_{g^\prime}}\vert i_1\cdots i_{g^{\prime}-1}\cdots i_{g}\rangle,
\end{equation}
where 
\begin{equation}\label{appeq:standardrep}
	\sum_{i_{g}=1}^{z_{g}}x^{i_1,\ldots,i_{g-1}}_{i_{g}} = 0.
\end{equation}

Since the constraint~\eqref{appeq:standardrep} is true for all values of the indices not summed over, the dimension of the irrep is straightforward to calculate and is given by  $\dim(g^\prime, g) = (n_{g^\prime}-1)\prod_{j=1}^{g^\prime-1}n_j$ for all values of $g$. Summing over the numbers of dimensions for all irreps (including the trivial ones from Eq.~\eqref{appeq:trivialsub}) residing in a single generation then gives us
\begin{equation}
	\sum_{g^\prime=0}^{g}\dim(g^\prime, g) = \prod_{j=1}^{g-1}n_j,
\end{equation}
which equals the number of sites in generation $g$. This is true for all generations and we can conclude that these irreps together cover the entire Hilbert space.

As we saw with the trivial irreps, where the hopping terms in the Hamiltonian mapped the irrep $(0, g)$ to $(0,g+1)$ and $(0,g-1)$ (except for $(0,0)$ which only maps to $(0, 1)$, since $(0,-1)$ does not exist), the hopping terms map the representation $(g^\prime, g)$ to $(g^\prime, g+1)$ and $(g^\prime, g-1)$ except for $g^\prime = g$ where the backwards hopping leads to
\begin{equation}
	\begin{split}
		\hat T^\dag \vert x_{g}, (g, g)\rangle &= \sum_{i_1,\ldots,i_{g}}x^{i_1,\ldots,i_{g-1}}_{i_{g}} \hat T^\dag\vert i_1\cdots i_{g-1}i_{g}\rangle\\
		&= \sum_{i_1,\ldots,i_{g}}x^{i_1,\ldots,i_{g-1}}_{i_{g}} \hat t^\dag_g\vert i_1\cdots i_{g-1}\rangle\\
		&=0,
	\end{split}
\end{equation}
where the last line follows from Eq.~\eqref{appeq:standardrep}. The conclusion we can draw from this is that we can decompose our Hamiltonian into chains; if we fix $g^\prime$, and use the same basis for each irrep $(g^\prime, g)$, we have $\dim(g^\prime, g^\prime)$ chains that decouple from the rest of the Hamiltonian and each other. They start at generation $g^\prime$ and extend up to $G$ (see Fig.~\ref{fig:decoupling} for an explicit example).

Let us now turn to the explicit forms of the Schr\"odinger equations for the linear chains. To this end, we begin by picking a basis in irrep $(g^\prime, g^\prime)$. We write the basis vectors in the form
\begin{equation}
	\vert m, (g^\prime, g^\prime) \rangle = \sum_{i_1,\ldots,i_{g^\prime}} b^{i_1,\ldots,i_{g^\prime-1}}_{m;i_{g^\prime}} \vert i_1\cdots i_{g^\prime} \rangle,
\end{equation}
where $m$ goes from $1$ to $\dim (g^\prime, g^\prime)$. By definition, we have $\langle n, (g^\prime, g^\prime)\vert m, (g^\prime, g^\prime)\rangle = \delta_{n,m}$. Each basis vector represents the edge of one chain. The subsequent sites along the chains are then the corresponding basis vectors in the irreps $(g^\prime, g)$:
\begin{equation}\label{appeq:norm}
	\vert m, (g^\prime, g) \rangle = \sum_{i_1,\ldots,i_{g}}\frac{b^{i_1,\ldots,i_{g^\prime-1}}_{m;i_{g^\prime}} \vert i_1\cdots i_{g^\prime}\cdots i_{g} \rangle}{\sqrt{\prod_{j=g^\prime + 1}^{g}n_j}} ,
\end{equation}
where the denominator is introduced in order to preserve normalization as we are summing over more and more sites when we increase $g$.

The normalization factor in Eq.~\eqref{appeq:norm} is important when considering the action of the Hamiltonian on these states. The forward hopping gives us
\begin{equation}
	\hat T\vert m, (g^\prime, g) \rangle = \sqrt{n_{g+1}}\hat t_{g}\vert m, (g^\prime, g+1) \rangle,
\end{equation}
and similarly we have for the backwards hopping
\begin{equation}
	\begin{split}
		\hat T^\dag \vert m, (g^\prime, g) \rangle &= \sum_{i_1,\ldots,i_{g}}\frac{b^{i_1,\ldots,i_{g^\prime-1}}_{m;i_{g^\prime}} \hat t_{g-1}^\dag\vert i_0\cdots i_{g^\prime}\cdots i_{g-1} \rangle}{\sqrt{\prod_{j=g^\prime + 1}^{g}n_j}}\\
		&= \sqrt{n_g}\hat t^\dag_{g-1}\vert m, (g^\prime, g-1) \rangle,
	\end{split}
\end{equation}
except, of course, for $g=g^\prime$. The above results combined with the onsite terms give us the equations for the linear chains as mentioned in the main text. 

\subsection{Tree Symmetry Subgroups}

By introducing intra-generation hopping terms between sites of the same branch, the symmetry can reduce from the full tree symmetry group to some proper subgroup of it. Fortunately, as we shall see below, the full symmetry of the symmetric groups $S_N$ are not necessary in order to decouple the system into linear chains.

From Cayley's theorem, we know that any finite-dimensional group is isomorphic to some subgroup of a symmetric group. This means that the elements of any group we consider will correspond to a permutation. Since a group consisting of permutations will always have a trivial irrep given by the vectors with all elements identical, it follows that the rest of the irreps will live in the subspace where all coordinates sum to zero. This property is important to ensure that the chains start at different generations, lest they will all couple together.

We must also restrict the valid groups to those that are not direct sums of two other groups. Such a group structure implies either of two things: the hopping from generation $g$ to $g+1$ is dependent on the index $i_{g+1}$ which means that it no longer maps representation $(g^\prime, g)$ to $(g^\prime, g+1)$, or the onsite term is index dependent in which case the onsite term does not trivially map a representation $(g^\prime, g)$ to itself anymore. In both cases, the result is that chains beginning at generations below $g$ will fork off into two or more chains. The extreme example of this is of course when the symmetry breaks down to only the trivial group containing only the identity element. In practice, there is then no symmetry and we are left with the original tree itself with no reductions.

To be more concrete, let us consider two examples. First, the cyclic group $C_N$ which is generated by a single group element $g$ such that $g^N$ is equal to the identity. The irreps of the cyclic group on complex vector spaces are all one-dimensional. In our case, $g$ will correspond to the matrix which shifts all the elements of a vector in $\mathbb{C}^N$ in a cyclic fashion such that there are no smaller subcycles under repeated application of the matrix (i.e. for each element in the vector, $g$ shifts them up one position up except for the uppermost element, which goes to the lowermost position). The matrix is unitary and any power of this matrix is diagonalized by the same eigenvectors. These eigenvectors can then each be thought of as the basis vectors to separate one-dimensional irreps. So the $(N-1)$-dimensional irrep of $S_N$ breaks down into $N-1$ one-dimensional irreps in $C_N$. For our chains, the practical consequence is that we are no longer free to pick any basis within the $(N-1)$-dimensional representation, but must instead use Fourier-type basis vectors for the chains.

Our second example is chosen to highlight that there are valid groups which do not have $C_N$ as a subgroup. To this end, consider the Klein four-group which is generated by two commuting elements that both square to identity. It is isomorphic to a subgroup in $S_4$, where the generating elements can be written as $\sigma^x \otimes \sigma^0$ and $\sigma^0\otimes \sigma^x$ in terms of the Pauli matrix $\sigma^x$ and the $2\times 2$ identity $\sigma^0$. Since all elements commute, they can all be simultaneously diagonalized, with each shared eigenvector serving as the basis vector defining one irrep each. Since it does not have $C_4$ as its subgroup, it cannot be a Fourier basis.

\subsection{Generalized Symmetries and Isotypic Decomposition}\label{app:iso_dec}

In this section we generalize the tree models, as well as their decoupling, discussed in the main text to a more generic setting. To be explicit, a generic RQT Hamiltonian with $n\ge 1$ levels can be expressed recursively as
\begin{align}\label{eq:Hn}
    \mathcal{H}_{n} = r_{n}+\sum_{b_{n}=1}^{B_{n}}\left[\mathcal{H}_{n-1}^{(b_{n})} + l_{n,n-1}^{(b_{n})}\right] + f_{n},
\end{align}
where $r_{n}$ is the on-site term for the root, $b_{n} = 1, \dotsc, B_{n}$ is the index for its branches and $B_n$ is the branching factor, $l_{n,n-1}^{(b_{n})}$ is the link term between the root and its $b_{n}$-th branch node, and $f_{n}$ is the coupling term between the root's different branches. By definition, $\mathcal{H}_{1} = r_{1}$ is an isolated node with $l_{1,0} = f_{1} = 0$. As a special case of RQTs, when the branching factors $B_{i}=1$ identically, the tree is reduced to a one-dimensional chain. In general, there will be $C_{i} = \prod_{j=i+1}^{n} B_{j}$ nodes at level $1\le i<n$ and we define $C_{n} = 1$ and $C_{i>n} = 0$. Based on the idea of recursion and for later convenience, we will refer to a node at level $i$ as an $i$-root and a tree with $i$ levels as an $i$-tree.

Following the recursion, each $i$-root can be uniquely indexed by a set of branch indices $\beta_{i} \equiv \{b_{i+1},\dots, b_{n}\}$ with $1\le i\le n$ and $\beta_{n} = \{\} \equiv \emptyset$, such that Eq.~\eqref{eq:Hn} can be expanded as
\begin{align}\label{eq:Hn_expand}
    \mathcal{H}_{n} = \sum_{i=1}^{n} \sum_{\beta_{i}} \left[r_{i}^{(\beta_{i})} + l_{i+1,i}^{(\beta_{i})} + f_{i}^{(\beta_{i})} \right],
\end{align}
where $l_{i+1,i}^{(\beta_{i})}$ stands for the link between $\beta_{i}$ and its immediate ancestor $\bar{\beta}_{i}$ such that if $\beta_{i} = \{b_{i+1}, b_{i+2},\dots, b_{n}\}$ then $\bar{\beta}_{i} = \{b_{i+2},\dots, b_{n}\}$, and by definition $l_{n+1,n} = 0$. For example, a minimal model of RQTs is given by
\begin{subequations}\label{eq:model_min}
	\begin{align}
		&r_{i}^{(\beta_{i})} = \epsilon_{i}\ket{\beta_{i}}\bra{\beta_{i}},\quad f_{i}^{(\beta_{i})} = 0, \\
		&l_{i+1,i}^{(\beta_{i})} = \tau_{i+1}\ket{\bar{\beta}_{i}}\bra{\beta_{i}} + \mathrm{h.c.},
	\end{align}
\end{subequations}
where $\ket{\beta_{i}}$ is the basis state representing the local degree of freedom at node $\beta_{i}$. More generally, we can take into account multiple inner degrees of freedom, labeled by $\alpha_{i}$ for an $i$-root, such that the basis of a RQT can be chosen as $\{\ket{\alpha_{i},\beta_{i}}\}$.

An RQT can be solved conveniently when there exists a hierarchy of symmetries associated with the tree structure, which we will refer to as the tree symmetry group. Similar to the Hamiltonian in Eq.~\eqref{eq:Hn}, the tree symmetry group extends (up to isomorphism) recursively by direct product:
\begin{align}
	\mathcal{G}_n = G_{n}\otimes\mathcal{G}_{n-1}.
\end{align}
Here $\mathcal{G}_{i}$ is the tree symmetry group for an $i$-tree, and $G_{i>1}$ is the symmetry group specifically introduced to the branches of an $i$-root. Explicitly, if $g_{i}\in G_{i}$, then the action of $g_{i}$ on a basis state is given by
\begin{align}
	&g_{i>j}\ket{\alpha_{j},\beta_{j}} \equiv g_{i}\ket{\alpha_{j},b_{j+1},\dots, b_{i}, \dots, b_{n}} \nonumber \\
	&\;= \sum_{b'_{i}}U(g_{i})_{b_{i} b'_{i}}\ket{\alpha_{j},b_{j+1},\dots, b'_{i}, \dots, b_{n}},
\end{align}
and $g_{i\le j}\ket{\alpha_{j},\beta_{j}} = \ket{\alpha_{j},\beta_{j}}$, where $U(g_{i})$ is the unitary transformation matrix that represents $g_{i}$. For the minimal model in Eq.~\eqref{eq:model_min}, $G_{i}$ is the permutation group of degree $B_{i}$, namely $G_{i} \cong S_{B_{i}}$, and each element of $G_{i}$ is a specific rearrangement of the $B_{i}$ branches of each $i$-root.

The strategy to solve an RQT can be demonstrated by using the minimal model as an example. To this end we write Eq.~\eqref{eq:model_min} in its recursive form
\begin{align}\label{eq:model_min_rec}
	\mathcal{H}_{n} = \epsilon_{n}\ket{\emptyset}\bra{\emptyset} + \sum_{b_{n}=1}^{B_{n}}\left[\left(\tau_{n}\ket{\emptyset}\bra{b_n} + \mathrm{h.c.}\right)+\mathcal{H}_{n-1}^{(b_{n})}\right],
\end{align}
where for transparency we denote $\ket{\emptyset} = \ket{\beta_{n}}$ and $\ket{b_{n}} = \ket{\beta_{n-1}}$. Since the cyclic group of order $B_{n}$ is a subgroup of the permutation group $S_{B_{n}}$, we can perform a partial Fourier transformation with respect to the index $b_{n}$ by defining
\begin{align}
	\ket{\dots, q_{n}} = \sum_{b_{n}=1}^{B_{n}}\frac{e^{\ii 2\pi q_{n}b_{n}/B_{n}}}{\sqrt{B_{n}}}\ket{\dots, b_{n}},
\end{align}
where by convention $q_{n}$ takes integers from $0$ to $B_{n}-1$, and $\dots$ in the kets stands for all possible indices, except for $b_{n}$, to specify a state in the original basis and they remain intact in the transformation. It follows that Eq.~\eqref{eq:model_min_rec} becomes
\begin{align}
	\mathcal{H}_{n} = &\,\epsilon_{n}\ket{\emptyset}\bra{\emptyset} + \left(\sqrt{B_{n}}\tau_{n}\ket{\emptyset}\bra{q_n=0} + \mathrm{h.c.}\right) \nonumber\\
	&+\sum_{q_{n}=0}^{B_{n}-1}\mathcal{H}_{n-1}^{(q_{n})},
\end{align}
where $\mathcal{H}_{n-1}^{(q_{n})}$ is formally identical to $\mathcal{H}_{n-1}^{(b_{n})}$ except that it is in the partially Fourier transformed basis. The key now is to notice that, unlike in the original basis, the sectors indexed by different $q_{n}$ are completely decoupled from each other with only $\mathcal{H}_{n-1}^{(q_{n}=0)}$ still connected to the $n$-root state $\ket{\emptyset}$. Clearly this transform-decouple procedure can be carried on to all the rest branch indices, such that a complete decomposition of the tree into one-dimensional chains will be achieved in the fully transformed Fourier basis $\{\ket{\kappa_{i} \equiv \{q_{i+1}, \dots, q_{n}\}}\}$. Moreover, two different basis states $\ket{\kappa_{i}}$ and $\ket{\kappa_{j}}$ belong to the same chain iif $i\ne j$ and, assuming $i<j$, for all $l>j$, $q_{l}(\kappa_{i})=q_{l}(\kappa_{j})$; otherwise $q_{l}(\kappa_{i})=0$. This immediately implies that the total number of chains is $C_{1}$, and among these chains $C_{i}-C_{i+1}$ start from $i$-roots (in the $\kappa$-representations) and hence are of length $i$. It is easy to check that the total number of nodes in all the chains is equal to that in the original tree, $\sum_{i=1}^{n} i(C_{i}-C_{i+1}) = \sum_{i=1}^{n} C_{i}$, such that the decomposition is indeed complete.

From a general point of view, the above chain decomposition is an example of the \textit{isotypic decomposition} of the Hilbert space according to the representations of the tree symmetry group $\mathcal{G}_n$. In this language, each chain is uniquely associated with a set of $\mathcal{G}_n$-isomorphic irreducible representation spaces, and each node in the chain corresponds to an irreducible representation space contained in the Hilbert space of a specific level of the tree. In the example of the minimal model, $\kappa_{i} = \{q_{i+1}, \dots, q_{n}\}$ is precisely a label for an irrep of $\mathcal{G}_n$ realized on the Hilbert space of $i$-roots, and the condition for $\ket{\kappa_{i}}$ and $\ket{\kappa_{j}}$ to belong to the same chain is precisely $\mathcal{G}_n$-isomorphism. In more general settings and with minor abuse of notation, the chain decomposition can be formally expressed as
\begin{align}\label{eq:decomp_gen}
	\mathcal{H} = \sum_{\kappa}\sum_{\kappa_{i}\cong \kappa_{j}\cong \kappa}\mathcal{H}^{(\kappa)}_{\kappa_{i}\kappa_{j}},
\end{align}
where the sum over $\kappa$ runs over all inequivalent irreps of $\mathcal{G}_n$ and each $\kappa$ labels an effective chain derived from an RQT.

\subsection{An RQT without a Root}\label{app:rootless}

\begin{figure}
	\includegraphics[width=0.75\linewidth]{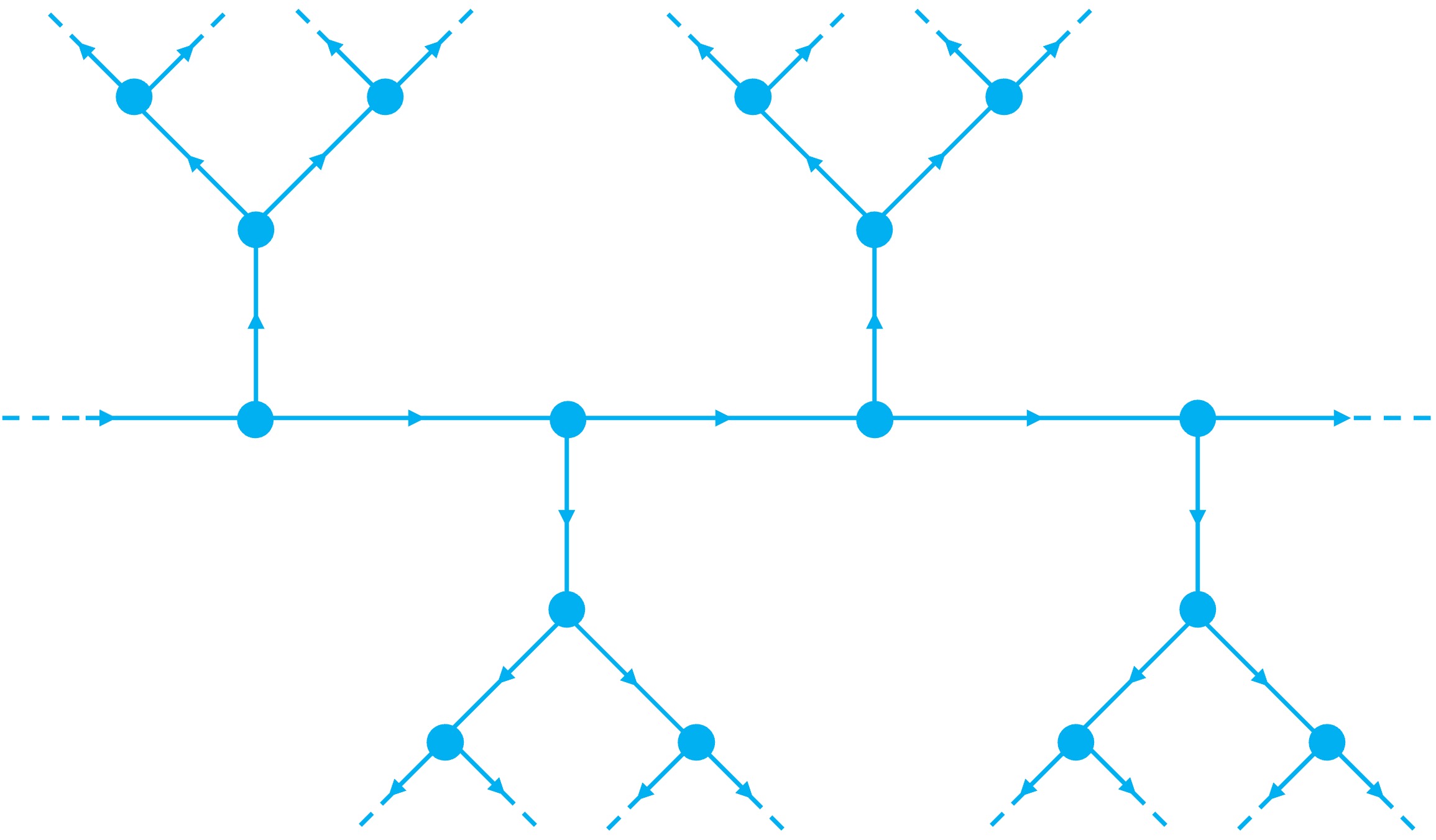}
	\caption{A presentation of a rooted tree without root.}
	\label{fig:rootless}
\end{figure}

Even though the RQTs of the main text have always and seemingly by definition come equipped with a root node, their existence is not inevitable: as a concrete example, we can take an infinite chain to which we attach $z-1$ semi-infinite RQTs (coordination number $z$) at each site. This creates a rootless RQT with coordination number $z$. An illustration of what such an infinite RQT might look like is provided in Fig.~\ref{fig:rootless}.

For clean rootless RQTs, the connection to the rooted RQTs is even more explicit: the translation invariance along the infinite chain allows us to perform a Fourier transformation, which results an RQT with a momentum-dependent root node. This affects neither the topology nor the bulk energies.

\section{Electric Circuit Simulation}\label{app:topcirc}

In this section, we elaborate the theory and the experiment for our electric circuit simulation of a Su-Schrieffer-Heeger Tree.

We first give a brief introduction to topolectrical circuits in order for this manuscript to be self-contained. For a more in-depth look at topolectrical circuits, please refer to \cite{lee:2018}. A topolectrical circuit is an electrical circuit where the electrical elements are chosen and arranged such that the equations governing the circuit are formally equivalent to the Schr\"odinger equation of a topological quantum system. The method provides measurable wave-number-resolved Laplacian eigenmodes forming an admittance band structure for the circuit which corresponds to the energy band structure of the quantum system. More critically, topological boundary resonances, corresponding to topological boundary modes in the quantum system, are present in impedance measurements, serving as robust signals for the presence of topological bands \cite{lee:2018,imhof:2018, bao:2019}.

The behavior of an RLC circuit is governed by its circuit Laplacian $J$, which is analogous to the Hamiltonian describing the energetics of a physical system. For an electrical circuit with $N$ nodes, define $I_j$ and $V_j$ as the current
entering the circuit at node $j$ and the potential difference between node $j$ and
the ground, respectively (node $j=0$ is taken to be the ground). If $g_{jl}$ is the conductance between node $j$ and $l$,
then Kirchhoff's law implies:
\begin{equation}
	\begin{split}
		I_j &= \sum_{l=0}^{N} g_{jl}\left(V_j - V_l\right)\\
		&= \sum_{l=1}^{N}\left[\sum_{m=1}^{N}g_{jm}\delta_{jl} + g_{j0}\delta_{jl} - g_{jl}\right]V_l = \sum_{l=1}^{N} J_{jl}V_l.
	\end{split}
\end{equation}
The matrix $J$ is called the grounded circuit Laplacian. Denoting the $n$th eigenmode of $J$ as $\psi_n$ and its eigenvalue as $j_n$, the inverse of $J$ can be expressed as
\begin{equation}
	G = \sum_{n}\frac{\psi_n \psi_n^\dag}{j_n},
\end{equation}
which we can use to express the two-point impedance  between node $a$ and $b$ in the form
\begin{equation}\label{appeq:z}
	\begin{split}
		Z_{ab} &= \sum_{i=a,b}\frac{G_{ai}I_i - G_{bi}I_i}{I}\\
		&= G_{aa} + G_{bb} - G_{ab} - G_{ba}\\
		&= \sum_{n=0}^{N} \frac{\Vert \psi_{n,a} - \psi_{n,b}\Vert^2}{j_n}.
	\end{split}
\end{equation}
The eigenvalue $j_n$ is a function of the frequency $\omega$ and the $n$th eigenenergy of the corresponding Hamiltonian. When the frequency is such that $j_n = 0$, we have a resonance which given Eq.~\eqref{appeq:z} tells us that we will expect a sharp increase in the impedance at those sites where the $n$th wavefunction has support.

Next we specify the experimental setup for our simulation. We employ printed circuit boards (PCBs) to realize the SSH tree. As shown in Fig.~\ref{fig:exp} (b), there are six generations of nodes including the root as the zeroth one. Each generation is connected to the previous and subsequent generation capacitively. The branching number alternates between generations with $n_{even} = 2$ and $n_{odd} = 3$. Again, to simulate a semi-infinite tree, a large imaginary term should be added to the diagonal terms of the terminal sites; correspondingly, we attach small resistors ($R_{\textrm{terminal}} \approx 0.2\Omega$) in parallel to the grounding of nodes on generation 5 as shown in box G of Fig.~\ref{fig:exp} (e).

Circuit elements on the PCB are arranged with sufficient spacing to reduce spurious mutual conductance. Moreover, the grounding is implemented as a layer of metal within the board instead of separated wires passing by circuit elements. One crucial point for designing a resonant circuit is to take into account the internal resistances of circuit elements, especially those of the small inductors. In our case, we adjust the ratio between the inductors and capacitors $L/C$ to adapt for the damping effect caused by the internal resistances ($\sim 0.3\Omega$) of the inductors, such that the system is underdamped and the resonances become detectable. In Fig.~\ref{fig:exp} (a) and (b), a photograph of the PCB used in our measurements is provided. The major parameters for the circuit elements are $L = 100\mu H$, $C=1nF$. In our numerical simulation of the experiment, shown in Fig.~\ref{fig:exp} (g), we have in addition adopted $R_{\textrm{terminal}} = 0.22\Omega$ and $R_C = 20\Omega$ for the internal (series) resistance of the capacitors.

Impedance measurements between pairs of nodes are taken by Keysight Impedance Analyzer E4990A. The analyzer serves as an AC current source, and sweeps through a range of frequencies while measuring impedance. At each sample frequency, several thousands of cycles are allowed to pass in order to stabilize the response of the circuit. Then the analyzer records the average impedance of about ten independent measurements at that frequency.

\bibstyle{apsrev4-1}
\bibliography{Treebib}

\end{document}